\documentclass[oneside,english]{amsart}
\usepackage[T1]{fontenc}
\usepackage[latin9]{inputenc}
\usepackage{amsthm}
\usepackage{amssymb}
\usepackage{graphicx}
\usepackage{esint}

\makeatletter
\numberwithin{equation}{section}
\numberwithin{figure}{section}
  \theoremstyle{remark}
  \newtheorem*{rem*}{\protect\remarkname}

\makeatother

\usepackage{babel}
  \providecommand{\remarkname}{Remark}

\begin{document}

\title{Projecting diffusion along the normal bundle of a plane curve}

\author{Carlos Valero Valdes\\
Departamento de Matematicas Aplicadas y Sistemas\\
Universidad Autonoma Metropolitana-Cuajimalpa\\
Mexico, D.F 01120, México\\
\\
Rafael Herrera Guzman\\
Centro de Investigacion en Matematicas (CIMAT)\\
Guanajuato, Gto\\
Mexico.}

\thanks{Partially supported by CONACyT grant 135106 }

\maketitle
\global\long\def\CC{\mathbb{C}}

\global\long\def\RR{\mathbb{R}}

\global\long\def\map{\rightarrow}

\global\long\def\mult{\mathcal{M}}

\global\long\def\EE{\mathcal{E}}

\global\long\def\OO{\mathcal{O}}

\global\long\def\FH{\mathcal{F}}

\global\long\def\CH{\mathcal{H}}

\global\long\def\VV{\mathcal{V}}

\global\long\def\PP{\mathcal{P}}

\global\long\def\CS{\mathcal{C}}

\global\long\def\tangent{T}

\global\long\def\cotangent{\tangent^{*}}

\global\long\def\conormal{\mathcal{C}}

\global\long\def\SS{\mathcal{S}}

\global\long\def\KK{\mathcal{K}}

\global\long\def\NN{\mathcal{N}}

\global\long\def\sym#1{\hbox{S}^{2}#1}

\global\long\def\symz#1{\hbox{S}_{0}^{2}#1}

\global\long\def\proj#1{\hbox{P}#1}

\global\long\def\SO{\hbox{SO}}

\global\long\def\GL{\hbox{GL}}

\global\long\def\U{\hbox{U}}

\global\long\def\tr{\hbox{tr}}

\global\long\def\ZZ{\mathbb{Z}}

\global\long\def\der#1#2{\frac{\partial#1}{\partial#2}}

\global\long\def\dder#1#2{\frac{\partial^{2}#1}{\partial#2^{2}}}

\global\long\def\covder{\hbox{D}}

\global\long\def\diff{d}

\global\long\def\dero#1{\frac{\partial}{\partial#1}}

\global\long\def\ind{\hbox{\, ind}}

\global\long\def\deg{\hbox{deg}}

\global\long\def\smb{S}

\global\long\def\DD{\mathcal{D}}

\global\long\def\QQ{\mathcal{Q}}

\global\long\def\RRR{\mathcal{R}}

\global\long\def\dd#1#2{\frac{d^{2}#1}{d#2^{2}}}

\global\long\def\d#1#2{\frac{d#1}{d#2}}

\global\long\def\and{\hbox{\,\, and\,\,\,}}

\global\long\def\where{\hbox{\,\, where\,\,\,}}

\global\long\def\KK{\mathcal{K}}

\global\long\def\II{\mathcal{I}}

\global\long\def\JJ{\mathcal{J}}

\global\long\def\Re{\hbox{Re}}

\global\long\def\Im{\hbox{Im}}

\begin{abstract}
The purpose of this paper is to provide new formulas for the effective
diffusion coefficient of a generalized Fick-Jacob's equation obtained
by projecting the two-dimensional diffusion equation along the normal
directions of an arbitrary curve on the plane. 
\end{abstract}

\section{Introduction}

\begin{figure}
\includegraphics[scale=0.4]{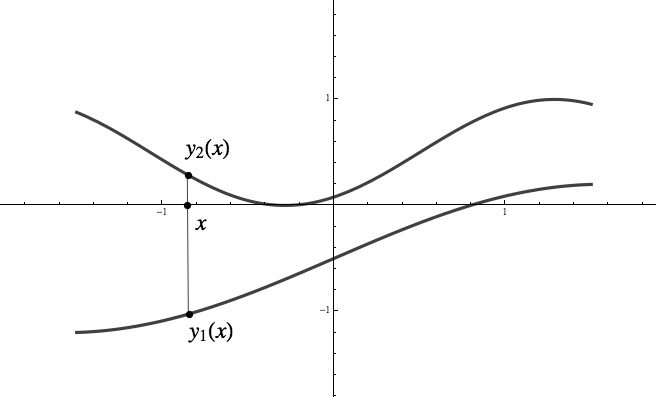}\caption{\label{fig:ChannelK0}Channel geometry defined as the set for point
$(x,y)$ such that $y_{1}(x)\leq y\leq y_{2}(x)$.}

\end{figure}

The problem of understanding spatially constrained diffusion in quasi-one
dimensional systems plays a fundamental role in diverse scientific
areas such as biology (e.g. channels in biological systems), chemistry
(e.g. pores in zeolites) and nano-technology (e.g. carbon nano-tubes).
Solving the diffusion equation for general channel-like constraining
geometries represents a very difficult task. One approach to overcome
this situation consists in reducing the degrees of freedom of the
problem by considering only the main direction of transport. More
concretely, consider a channel-like geometry bounded from below and
above by two functions $y=y_{1}(x)$ and $y=y_{2}(x)$ (see Figure
\ref{fig:ChannelK0}), and a concentration density function $P=P(x,y,t)$
satisfying the diffusion equation 
\[
\der Pt(x,y,t)=D_{0}\left(\dder Px(x,y,t)+\dder Py(x,y,t)\right)
\]
and having no flow across the channel's walls.
If we integrate $P$ with respect to the $y$-coordinate we obtain
an effective density function
\[
p(x,t)=\int_{y_{1}(x)}^{y_{2}(x)}P(x,y,t)dy,
\]
whose values represent the total concentration density along vertical
cross sections of the channel. The dynamics of $p$ can be modeled
in an approximate manner by a generalized Fick-Jacob's equation 
\[
\der pt(x,t)=\dero x\left(\DD(x)w(x)\dero x\left(\frac{p(x,t)}{w(x)}\right)\right),
\]
where the function $\DD=\DD(x)$ is known as the effective diffusion
coefficient and $w(x)=y_{2}(x)-y_{1}(x)$ is  the channel's width.
The precision of the approximation that the generalized Fick-Jacob's
provide compared to the true time evolution of $p$ depends on an
adequate estimation of the effective diffusion coefficient. The simplest
estimate for $\DD$ is the one given by $\DD(x)=D_{0}$, which arises
from the assumption that the density $P$ stabilizes infinitely fast
in the transversal direction (i.e $P$ is independent of $y$). Better
estimations of $\DD$ must account for the fact that in reality the
transversal diffusion rate is finite. Work in this direction was carried
out by Zwanzig in \cite{kn:entropybarrierzwanzig} where he obtained
the formula 
\[
\DD(x)=D_{0}\left(\frac{1}{1+w'(x){}^{2}/12}\right)\where w'(x)=\d wx(x),
\]
which then Bradley generalized in \cite{kn:bradley} to 
\[
\DD(x)=D_{0}\left(\frac{1}{1+y'_{0}(x)^{2}+w'(x)^{2}/12}\right)\where y_{0}(x)=\frac{y_{1}(x)+y_{2}(x)}{2},
\]
to deal with non-symmetric channels (i.e $y_{0}\not=0)$. Reguera
and Rubí in \cite{kn:reguerarubi} also proposed an improvement of
Zwanzig's formula, given by

\[
\DD(x)=D_{0}\left(\frac{1}{1+w'(x){}^{2}/4}\right)^{1/3}.
\]
Later on Kalinay \& Percus developed in \cite{kn:kp-diffusion-projection}
a systematic procedure to obtain increasingly better formulas to $\DD$
, and which they used to derived the formula

\[
\DD(x)=D_{0}\left(\frac{\arctan(w'(x)/2)}{w'(x)/2}\right)
\]
under the assumption that $y_{1}(x)=0$. This formula was then extended,
dropping the assumption $y_{1}(x)=0$, by L. Dagdug \& I. Pineda in
\cite{kn:di-projection-diffusion} to obtain 

\[
\DD(x)=D_{0}\left(\frac{\arctan(y'_{0}(x)+w'(x)/2)-\arctan(y'_{0}(x)-w'(x)/2)}{w'(x)/2}\right).
\]
\begin{figure}
\includegraphics[scale=0.45]{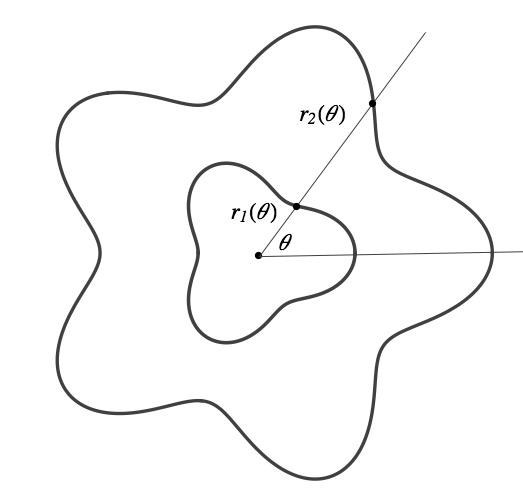}\caption{\label{fig:PolarCoordChannel}Channel in polar coordinates}

\end{figure}

The results described above deal with the ``projection'' of a two
dimensional density function onto a the $x$-coordinate of the standard
coordinates $(x,y)$ on the plane. As an example where these coordinates
are not the best choice consider a channel-like geometry as the one
depicted in Figure \ref{fig:PolarCoordChannel}, in which it is more
natural to use polar coordinates. The use of coordinate functions
that follow in a ``natural manner'' the geometry of the channel
can help in the calculation of effective diffusion coefficient (see
\cite{kn:kp-Extendex-fj-variational}). Furthermore, to study quasi-one
dimensional diffusion on a curved surface (e.g the surface of a cell)
one is forced from the beginning to consider more general coordinate
systems than the standard $(x,y$) system in the plane.

\begin{figure}
\includegraphics[scale=0.4]{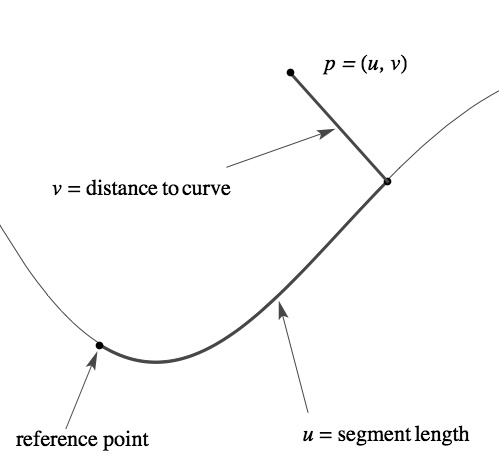}\caption{\label{fig:UVCoordinates}Point $p$ in the plane with $(u,v)$ coordinates}

\end{figure}

The purpose of these paper is to generalize the work describe above
by deriving formulas for the effective diffusion coefficient (in the
infinite and finite transversal diffusion rate cases) for channels
defined on a system of coordinates $u,v$ constructed as follows.
For an arbitrary curve in the plane (which we will refer as the base
curve) we will let $u$ be base curve's arc-length parameter with
respect to a fixed reference point, and we will let $v$ be the normal
distance to the base curve (see Figure \ref{fig:UVCoordinates}).
As before, we can then construct channels over the base curve by using
a pair of functions $v_{1}=v_{1}(u)$ and $v_{2}=v_{2}(u)$ (see Figure
\ref{fig:Channel-over-normal}). In a similar way as in the case of
rectangular coordinates, from the density function $P$ we can construct
an effective density function $p=p(u,t)$ whose evolution can be modeled
(in an approximate way) by a generalized Fick-Jacob's equation of
the form
\[
\der pt(u,t)=\dero u\left(\DD(u)\sigma(u)\dero u\left(\frac{p(u,t)}{\sigma(u)}\right)\right).
\]
In the above formula $\DD=\DD(u)$ stands for the effective diffusion
coefficient, and the function $\sigma$ is given by
\[
\sigma(u)=\d Au(u),
\]
where $A(u)$ defined as the area of the channel from the transversal
section at the reference point to the transversal section over the
point on the base curve with coordinate $u$. In the case when the
base curve in the $x$-axis the function $\sigma$ coincides with
the channel's width function. 

The outline of our article is as follows:
\begin{itemize}
\item In section 2 we will show that if the concentration density in the
plane satisfies the continuity equation (given by formula \ref{eq:Continuity Equation})
then the corresponding effective density also obeys a continuity equation
(see formula \ref{eq:reducedContinuity}). This last equation, which
we will refer as the effective continuity equation, will serve as
the basis for the work that follows in the rest of the article.
\item In section 3 we will derive the formula for the effective diffusion
coefficient corresponding to the infinite transversal diffusion rate
(see formula \ref{eq:DZerothOrder}). The obtained formula generalizes
the standard Fick-Jacob's equation in the sense it is a particular
case of ours for a base curve that has zero curvature (i.e it is a
straight line). We end the section by applying our formula to derive
some interesting physical properties of symmetric channels of constant
width.
\item In sub-section 4.1 we derive our first formula for the effective diffusion
coefficient corresponding to a finite transversal diffusion rate.
Our formula in this case involves tangential information (i.e derivatives)
of the curves forming the upper and lower boundaries of the channel.
We show that the obtained formula generalizes the results of L. Dagdug
and I. Pineda \cite{kn:di-projection-diffusion}. Again, their formula
can be recovered from ours as the case when the base curve has zero
curvature. It is interesting to observe that in the zero curvature
case the effective diffusion coefficient is always less than or equal
to the diffusion coefficient $D_{0}$ for the corresponding the 2-dimensional
diffusion equation. In contrast, in the general case of projecting
onto an arbitrary curve the effective diffusion coefficient can be
less than, greater than or equal to $D_{0}$.
\item In sub-section 4.2 we derive our second formula for the effective
diffusion coefficient corresponding to a finite transversal diffusion
rate. In this case our formula involves both tangential and curvature
information of the curves forming of the upper and lower boundaries
of the channel. We show that for symmetric channels of constant width
the infinite transversal diffusion rate case coincides with current
case. This confirms our intuition that the choice of an appropriate
coordinate systems simplifies the computation of the effective diffusion
coefficient.
\end{itemize}
We will use the language of the differential geometry of planar curves
to write down the formula for $\DD$ in a coordinate-free manner.
Furthermore, we use complex analysis to provide a more compact expression
for $\DD$ than would otherwise be obtained by using only real analysis.
In particular, our use of the complex logarithm function will help
us unify different cases which otherwise would appear to be unrelated.

\section{The effective continuity equation on the plane}

We will identify 2-dimensional space $\RR^{2}$ with the complex plane
$\CC$ by means of the correspondence: $(x,y)\leftrightarrow z=x+iy$.
To distinguish the dot product in $\RR^{2}$ from the complex product
in $\CC$ we will use the symbol $<,>$ for the former and simple
juxtaposition for the latter.

\begin{figure}
\includegraphics[scale=0.5]{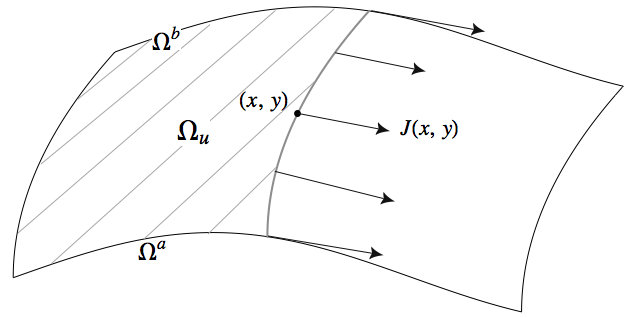}\caption{\label{fig:Density-flow}Density flow along region $\Omega$.}
\end{figure}

We are interested in describing transport processes on a region $\Omega$
of the plane which are modeled by the continuity equation 
\begin{equation}
\der Pt+\hbox{div}(J)=0,\label{eq:Continuity Equation}
\end{equation}
where $P=P(z,t)$ is a real valued density function and $J=J(z,t)$
is its corresponding complex valued flux. We will apply a dimensionality
reduction technique to this equation as follows. Let us assume that
$\Omega$ is parametrized by a smooth map $\varphi:[u_{1},u_{2}]\times[a,b]\rightarrow\Omega$,
and consider the following subsets of $\Omega$ (see Figure \ref{fig:Density-flow})
\begin{eqnarray*}
\Omega_{u} & = & \varphi([u_{1},u]\times[a,b]),\\
\Omega^{a} & = & \{\varphi(u,a)|u_{1}\leq u\leq u_{2}\},\\
\Omega^{b} & = & \{\varphi(u,b)|u_{1}\leq u\leq u_{2}\}.
\end{eqnarray*}
We will refer to $\Omega^{a}$ and $\Omega^{b}$ as lower and upper
walls of $\Omega$. The total concentration of $P$ in the region
$\Omega_{u}$ is given by
\[
C(u,t)=\int_{a}^{b}\int_{u_{1}}^{u}P(\varphi(s,v),t)\det(\varphi'(s,v))dsdv,
\]
where $\varphi'$ is the Jacobian matrix of $\varphi$. The \emph{effective
density} $p$ associated to $P$ is defined as 
\begin{equation}
p(u,t)=\d Cu(u,t)=\int_{a}^{b}P(\varphi(u,v),t)\det(\varphi'(u,v))dv,\label{eq:reducedDensity}
\end{equation}
and the corresponding \emph{effective flux} $j$ by
\[
j(u,t)=\int_{a}^{b}\left\langle iJ(\varphi(u,v),t),\der{\varphi}v(u,v)\right\rangle dv,
\]
The quantity $p(u,t)$ measures the density concentration at time
$t$ along the cross section parametrized by $v\mapsto\varphi(u,v)$,
and $j(u,t)$ measures the flux density along the same cross section.
If we assume that there is no flux of $P$ across the upper and lower
walls of $\Omega$ then the two-dimensional continuity equation (\ref{eq:Continuity Equation})
implies the effective continuity equation

\begin{equation}
\der pt(u,t)+\der ju(u,t)=0.\label{eq:reducedContinuity}
\end{equation}
If we impose Fick's law
\[
J=-D_{0}\nabla P,
\]
for a constant diffusion coefficient $D_{0}$, the two-dimensional
continuity equation becomes \emph{the diffusion equation} 
\[
\der Pt=D_{0}\Delta P.
\]
 In this case, the effective flux becomes

\begin{equation}
j(u,t)=-D_{0}\int_{a}^{b}\left\langle i\nabla P(\varphi(u,v),t),\der{\varphi}v(u,v)\right\rangle dv.\label{eq:reducedFluxFormula}
\end{equation}

\section{A generalized Fick-Jacob's  equation on the normal bundle of a plane
curve : infinite transversal diffusion rate case }

\begin{figure}
\includegraphics[scale=0.55]{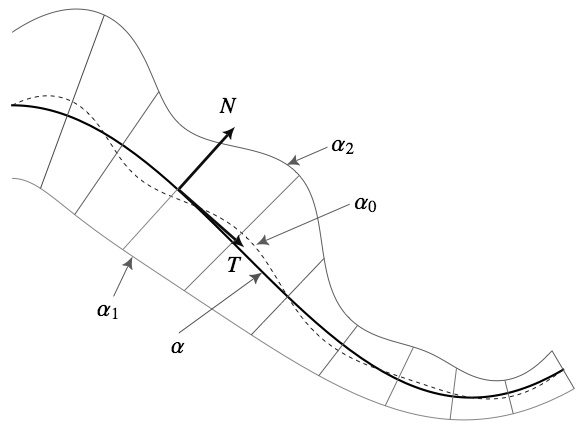}\caption{\label{fig:Channel-over-normal}Channel over normal bundle of a curve}
\end{figure}
\begin{figure}
\includegraphics[scale=0.5]{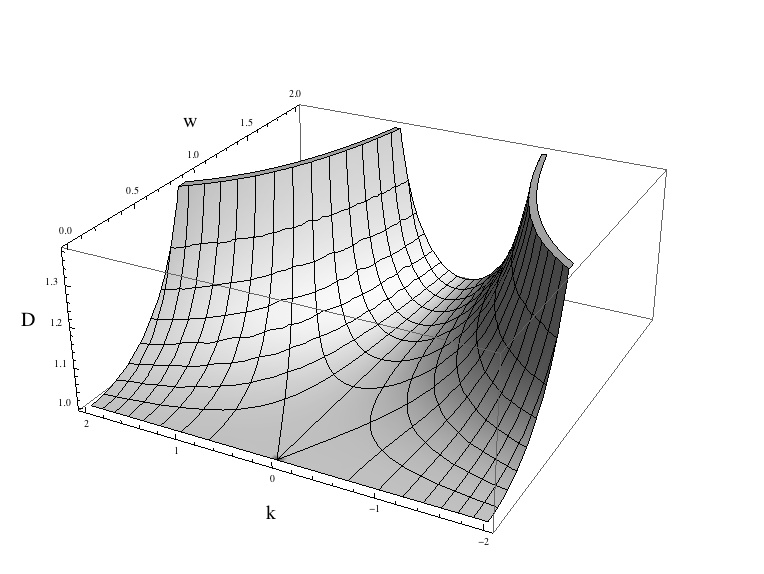}\caption{\label{fig:Graphic-of-D}Graphic of $\DD$ in the on the $(k,w)$-plane
for $D_{0}=1$.}
\end{figure}

Consider a curve $\alpha:[u_{1},u_{2}]\rightarrow\CC$ parametrized
by the arc-length parameter $u$. Under this assumption the vectors
\[
T=\der{\alpha}u\and N=i\der{\alpha}u
\]
are the unit tangent and unit normal vectors of $\alpha$. We will
consider a region $\Omega$ parametrized by the map $\varphi:[u_{1},u_{2}]\times[-1,1]\rightarrow\RR^{2}$
given by 
\begin{equation}
\varphi(u,v)=\alpha(u)+s(u,v)N(u),\label{eq:NormalBundleParametrisation}
\end{equation}
where
\[
s(u,v)=v_{0}(u)+vw(u)/2,
\]
for given functions $v_{0},w:[u_{1},u_{2}]\rightarrow\RR$. We will
refer to $v_{0}$ as the \emph{middle curve function} and to $w$
as the \emph{width function}. The lower and upper curves $\alpha_{1}$
and $\alpha_{2}$ are then given by 
\begin{eqnarray}
\alpha_{1}(u) & = & \varphi(u,-1)=\alpha(u)+(v_{0}(u)-w(u)/2)N,\label{eq:alpha1}\\
\alpha_{2}(u) & = & \varphi(u,1)=\alpha(u)+(v_{0}(u)+w(u)/2)N,\label{eq:alpha2}
\end{eqnarray}
and we define the middle curve $\alpha_{0}$ by
\[
\alpha_{0}(u)=\varphi(u,0)=\alpha(u)+v_{0}(u)N.
\]
Observe that in general $\alpha_{0}$ and $\alpha$ do not coincide
(see Figure \ref{fig:Channel-over-normal}). By using the formulas
(see Appendix 1) 
\[
\d Tu=kN\and\d Nu=-kT,
\]
where $k$ is the curvature function of $\alpha$, we obtain the equations
\begin{eqnarray*}
\der{\varphi}u & = & (1-sk)T+\der suN,\\
\der{\varphi}v & = & (w/2)N,
\end{eqnarray*}
and
\begin{equation}
\det(\varphi')=(1-sk)(w/2).\label{eq:JacobianAndCurvature}
\end{equation}
In this case, the effective density (\ref{eq:reducedDensity}) becomes
\[
p=\int_{-1}^{1}P(1-sk)(w/2)dv.
\]
To compute the effective flux (\ref{eq:reducedFluxFormula}) observe
that 
\[
\der Pu=\left\langle \nabla P,\left(\der{\varphi}u\right)\right\rangle =(1-sk)\left\langle \nabla P,T\right\rangle +\der su\left\langle \nabla P,N\right\rangle 
\]
and
\[
\der Pv=\left\langle \nabla P,\left(\der{\varphi}v\right)\right\rangle =(w/2)\left\langle \nabla P,N\right\rangle .
\]
From the above formulas and the definition of $j$ we obtain
\begin{eqnarray*}
j & = & -D_{0}\int_{-1}^{1}(w/2)<\nabla P,T>dv\\
 & = & -D_{0}\int_{-1}^{1}(1-sk)^{-1}\left(\der Pu\frac{w}{2}-\der Pv\der su\right)dv.
\end{eqnarray*}
The hypothesis of an infinite diffusion rate along the $v$-variable
means that $P$ is independent of that variable, and hence we obtain
\begin{eqnarray*}
p(u,t) & = & \sigma(u)P(u,t),\\
j(u,t) & = & -D_{0}\gamma(u)\der Pu(u,t),
\end{eqnarray*}
where
\begin{eqnarray*}
\sigma(u) & = & \frac{1}{2}\int_{-1}^{1}(1-s(u,v)k(u))w(u)dv,\\
\gamma(u) & = & \frac{1}{2}\int_{-1}^{1}(1-s(u,v)k(u))^{-1}w(u)dv.
\end{eqnarray*}
From the results above, we conclude that the effective continuity
equation (\ref{eq:reducedContinuity}) becomes the following generalized
Fick-Jacob's equation 
\begin{equation}
\der pt(u,t)=\dero u\left(\DD(u)\sigma(u)\dero u\left(\frac{p(u,t)}{\sigma(u)}\right)\right),\label{eq:FickJacobsEqInCurveBundle}
\end{equation}
where
\begin{eqnarray}
\DD & = & \left(\frac{D_{0}}{kw}\right)\left(\frac{1}{1-kv_{0}}\right)\log\left(\frac{1-k(v_{0}-w/2)}{1-k(v_{0}+w/2)}\right),\label{eq:DZerothOrder}\\
\sigma & = & w(1-kv_{0}).\nonumber 
\end{eqnarray}
We will refer to $\DD$ as the \emph{effective diffusion coefficient}.
Observe that $\sigma$ in equation (\ref{eq:FickJacobsEqInCurveBundle})
plays the role of the width in the classical Fick-Jacob's equation.
The function $\sigma$ is an area density since the area of the region
$\Omega_{u}=\varphi([u_{1},u])$ is given by
\[
A(u)=\int_{u_{1}}^{u_{2}}\sigma(a)da.
\]
 For a symmetric channel ($v_{0}=0$) we have that $\sigma=w$, i.e.
we recover the width function, and
\begin{equation}
\DD=\left(\frac{D_{0}}{kw}\right)\log\left(\frac{1+kw/2}{1-kw/2}\right).\label{eq:SymmCstWidth}
\end{equation}
If we consider $\DD$ as a function of $w$ and $k$, its level sets
are the hyperbolas $kw=\hbox{constant}$ and its gradient lines are
the hyperbolas $w^{2}-k^{2}=\hbox{constant}$ (see Figure \ref{fig:Graphic-of-D}).
It is easily seen that
\begin{eqnarray*}
\lim_{kw\mapsto0}\DD(k,w) & = & \DD_{0},\\
\lim_{kw\mapsto2}\DD(k,w) & = & \infty.
\end{eqnarray*}
Thus we have the following observations about symmetric channels of
constant width.
\begin{enumerate}
\item The the value of the effective diffusion coefficient remains constant
if we keep $w$ and $k$ inversely proportional. In particular, if
we increase the curvature of a channel (e.g by bending it) then we
must reduce its width to keep the effective diffusion constant.
\item The effective diffusion coefficient coincides with the diffusion coefficient
of the non-reduced diffusion equation when either the curve has zero
width and arbitrary curvature function, or the base curve is a straight
line (i.e $k=0$) with arbitrary width function.
\item The limit $\lim_{kw\mapsto2}\DD(k,w)=\infty$ means the channel is
such that its upper or lower walls contain focal points of the base
curve. This situation is a consequence of the degeneracy of the coordinate
system induced by the normal lines of the base curve. To avoid this
complication the upper and lower walls of the channel must stay close
to the base curve, assumption which is consistent with the fact that
we are dealing with narrow channels.
\item Observe that for symmetric channels of constant width the effective
diffusion coefficient is always larger than $D_{0}$. For example,
if we take a straight cylindrical tube and bend it (so that its walls
remain at a constant distance from the base curve) then the effective
diffusion increases.\end{enumerate}
\begin{rem*}
As explained in point 2 above, the effective diffusion coefficient
can become infinite due to the existence of focal points of the base
curve. In the case of projecting diffusion onto a straight line this
does not occur since the only focal point is ``hidden'' at infinity.
Even if the effect of infinite effective diffusion can be avoided
when projecting diffusion onto a straight line, it might be unavoidable
in the case of projecting diffusion along a geodesic on surfaces like
the sphere. The mathematical explanation for this is that the sphere
is a compact surface, but the plane is not.
\end{rem*}

\section{Finite transversal diffusion rate cases}

In this section we will deduce two new formulas for the effective
diffusion coefficient that improve the estimation obtained in the
previous section. To do this we will apply a technique described by
Kalinay \& Percus in \cite{kn:kp-fick-jacob-correction}. The idea
consists in approximating the original channel by a simpler one, and
estimating the effective diffusion coefficient by finding an explicit
stable solution of the diffusion equation with the appropriate boundary
conditions on the simpler channel. We will incorporate this idea into
our work by using complex analytic functions as follows. 

Recall that real and imaginary parts of a complex analytic function
are stable solutions of the diffusion equation, i.e. they are harmonic
functions. Let $\Omega$ be the channel region and suppose that we
can find an explicit formula for a complex analytic function on $\PP:\Omega\rightarrow\CC$
whose real part has zero flux along the upper and lower walls of $\Omega$.
By the Cauchy-Riemann equations this condition is equivalent to the
imaginary part of $\PP$ being constant on the upper and lower walls
of $\Omega$. From the results of the previous section we have that
the effective density (\ref{eq:reducedDensity}) for the real part
of $\PP$ is the real part of the complex valued function 
\begin{equation}
\rho(u)=\int_{-1}^{1}\PP(\varphi(u,v))\left(1-\left(v_{0}(u)+vw(u)/2\right)k\right)(w(u)/2)dv.\label{eq:rhoIntegral}
\end{equation}
Since the real and imaginary parts of $\PP$ satisfy the Cauchy-Riemann
equations
\[
\nabla(\Im(\PP))=i\nabla(\Re(\PP)),
\]
we can compute the effective flux (\ref{eq:reducedFluxFormula}) of
$\Re(\PP)$ as follows
\begin{eqnarray*}
j(u) & = & -D_{0}\int_{-1}^{1}\left\langle i\nabla(\Re(\PP)),\der{\varphi}v(u,v)\right\rangle dv\\
 & = & -D_{0}\int_{-1}^{1}\left\langle \nabla(\Im(\PP)),\der{\varphi}v(u,v)\right\rangle dv\\
 & = & -D_{0}\Im(\PP(\varphi(u,1))-\PP(\varphi(u,-1))).
\end{eqnarray*}
If we equate this flux with the one in the generalized Fick-Jacob's
equation (\ref{eq:FickJacobsEqInCurveBundle}), i.e 
\begin{equation}
j(u)=-\DD(u)\sigma(u)\dero u\left(\frac{\Re(\rho(u))}{\sigma(u)}\right),\label{eq:FluxFromFJ}
\end{equation}
we obtain the following formula for the effective diffusion coefficient
\begin{equation}
\DD(u)=D_{0}\frac{\Im(\DD_{1}(u))}{\Re(\DD_{2}(u))},\label{eq:MasterFormulaForD}
\end{equation}
where
\begin{equation}
\DD_{1}(u)=\PP(\alpha_{2}(u))-\PP(\alpha_{1}(u))\label{eq:D1}
\end{equation}
and
\begin{equation}
\DD_{2}(u)=\sigma(u)\dero u\left(\rho(u)/\sigma(u)\right)\label{eq:D2}
\end{equation}
Equation (\ref{eq:MasterFormulaForD}) gives us an explicit formula
for $\DD$ if we can compute $\PP$ explicitly and calculate the integral
(\ref{eq:rhoIntegral}). Since finding $\PP$ or solving the integral
in \ref{eq:rhoIntegral} explicitly cannot be easily achieved in general,
we will estimate $\DD$ by computing $\PP$ for a secondary channel
that approximates locally the original one.
\begin{rem*}
The Riemann Mapping Theorem states that if $\Omega\subset\CC$ is
open and simply connected, then there exists a bi-holomorphism $F:\Omega\rightarrow A$,
where $A$ is the unit disk in $\CC$. If we knew such a map $F$
we could define $\PP(z)=\log(F(z))$.
\end{rem*}

\subsection{Linear Case}

\begin{figure}
\includegraphics[scale=0.35]{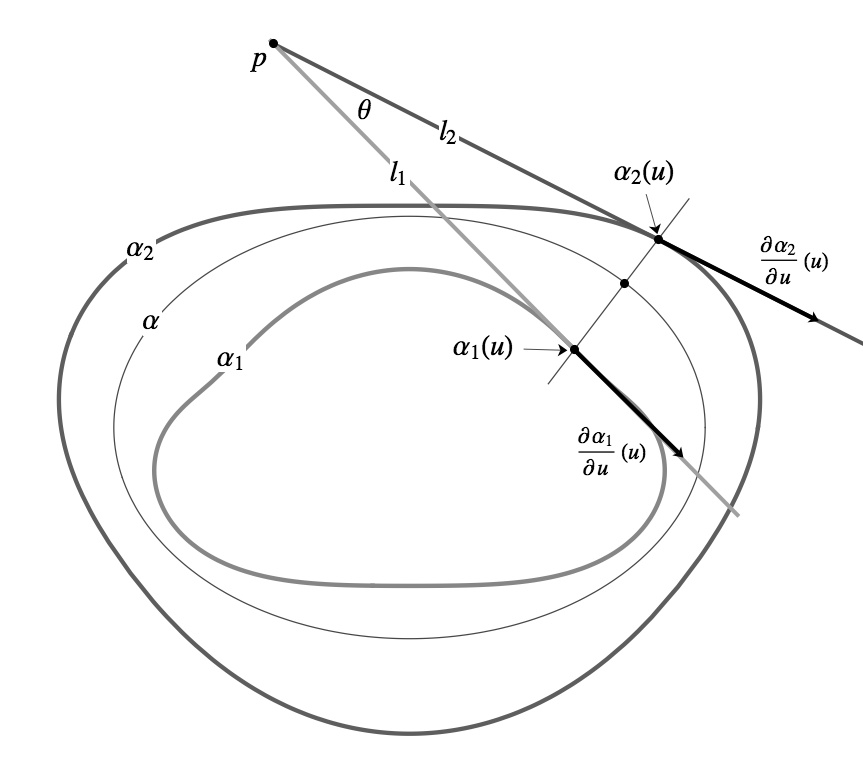}\caption{\label{fig:Linear-aproximation-to}Linear approximation to a channel}
\end{figure}

To simplify notation we will write 
\[
\alpha'_{1}=\der{\alpha_{1}}u,\alpha'_{2}=\der{\alpha_{2}}u,v_{0}'=\der{v_{0}}u\and w'=\der wu.
\]
For a fixed value of $u\in[u_{1},u_{2}]$ we will approximate the
channel $\Omega$ by a channel having lower and upper walls formed
by the tangent lines $l_{1}$ and $l_{2}$ to the curves $\alpha_{1}$
and $\alpha_{2}$ (see Figure \ref{fig:Linear-aproximation-to}).
If the lines $l_{1}$ and $l_{2}$ intersect at the point $p$ then
the imaginary part of the function
\[
\PP(z)=\log(z-p)=\log(|z-p|)+i\arg(z-p)
\]
is constant along $l_{1}$ and $l_{2}$, and hence its real part is
a harmonic function having zero flux on these lines.

\begin{rem*}
If we more generally we chose $\PP(z)=a\log(z-p)+b$ for $a,b\in\RR$,
our formula for $\DD$ turns out to be independent of $a,b$.
\end{rem*}
From the geometry of Figure \ref{fig:Linear-aproximation-to} we obtain
that
\begin{eqnarray}
\Im(\DD_{1}) & = & \arg(\alpha_{2}(u)-p)-\arg(\alpha_{1}(u)-p),\label{eq:D1Linear}\\
 & = & \arg\left(\alpha'_{2}(u)\right)-\arg\left(\alpha'_{1}(u)\right),
\end{eqnarray}
i.e the quantity $\Im(\DD_{1})$ is the angle $\theta$ between $l_{1}$
and $l_{2}$. To compute $\DD_{2}$ we need to compute $\rho$ explicitly.
To do this we will approximate the base curve $\alpha$ by its circle
of curvature at the point of interest, which amounts to assuming that
\begin{equation}
\alpha(u)=f-i\exp(iku)/k,\label{eq:baseCircle}
\end{equation}
so that 
\[
N(u)=i\exp(iku).
\]
By computing the integral (\ref{eq:rhoIntegral}) we obtain 
\[
\rho(u)=Q_{2}(u)\log(\alpha_{2}(u)-p)-Q_{1}(u)\log(\alpha_{1}(u)-p)-R(u),
\]
where 
\begin{eqnarray*}
R(u) & = & \left(\frac{k}{2N^{2}(u)}\right)(\alpha_{2}(u)-\alpha_{1}(u))(f-p-(\alpha_{0}(u)-f)),\\
Q_{1}(u) & = & \left(\frac{k}{2N^{2}(u)}\right)(\alpha_{1}(u)-p)(f-p-(\alpha_{1}(u)-f)),\\
Q_{2}(u) & = & \left(\frac{k}{2N^{2}(u)}\right)(\alpha_{2}(u)-p)(f-p-(\alpha_{2}(u)-f)).
\end{eqnarray*}
If we substitute the above formula for $\rho$ in equation (\ref{eq:D2})
we obtain 
\begin{equation}
\DD_{2}(u)=\RRR(u)+(\QQ_{2}(u)-\QQ_{1}(u))\log\left(\frac{\alpha_{2}(u)-p}{\alpha_{1}(u)-p}\right),\label{eq:D2Linear}
\end{equation}
where
\[
\RRR=\frac{k(\alpha_{0}-f-(f-p))\left((\alpha_{1}-f)\alpha'_{1}-(\alpha_{2}-f)\alpha'_{2}\right)}{2N^{2}(\alpha_{0}-f)}
\]
and
\begin{eqnarray*}
\QQ_{1} & = & \frac{k(f-p-(\alpha_{1}-f))(\alpha_{2}-f)(\alpha_{1}-p)\alpha'_{2}}{2N^{2}(\alpha_{2}-\alpha_{1})(\alpha_{0}-f)},\\
\QQ_{2} & = & \frac{k(f-p-(\alpha_{2}-f))(\alpha_{1}-f)(\alpha_{2}-p)\alpha'_{1}}{2N^{2}(\alpha_{2}-\alpha_{1})(\alpha_{0}-f)}.
\end{eqnarray*}
Having computed $\DD_{1}$ and $\DD_{2}$ explicitly we can compute
$\DD$ by using formula \ref{eq:MasterFormulaForD}. Observe that
our formulas for $\DD_{1}$ and $\DD_{2}$ involve purely geometric
quantities associated to the curves $\alpha_{1},\alpha_{2}$ and $\alpha$.

\subsubsection*{Recovering the case of projection onto a straight line}

We can recover the formula for the effective diffusion coefficient
obtained by L. Dagdug and I. Pineda \cite{kn:di-projection-diffusion}
as a particular case of ours in the limiting case when $k=0$, i.e.
by projecting diffusion onto a straight line. More precisely, observe
that the curve $\alpha$ defined by (\ref{eq:baseCircle}) above satisfies
\[
f=\frac{i}{k}\hbox{\,\, implies\,\,}\lim_{k\mapsto0}\alpha(u)=u,
\]
so that the straight line given by the $u$-axis can be seen as a
limiting case of the curve $\alpha$ when $k\mapsto0$. 

On the other hand, from the formulas for $\QQ_{1},\QQ_{2}$ and $\RRR$
above, and letting $f=i/k$, we can compute 
\begin{eqnarray*}
\lim_{k\mapsto0}\RRR & = & i(\alpha'_{1}-\alpha'_{2}),\\
\lim_{k\mapsto0}(\QQ_{2}-\QQ_{1}) & = & i\left(\frac{(\alpha_{2}-p)\alpha'_{1}-(\alpha_{1}-p)\alpha'_{2}}{\alpha_{1}-\alpha_{2}}\right).
\end{eqnarray*}
The velocity vectors $\alpha'_{1}$ and $\alpha'_{2}$ are connected
by the fact that they are obtained by sweeping the normal line of
$\alpha$ along the curve $\alpha_{1}$ and $\alpha_{2}$. This property
and the fact that for $k=0$ we have that $\alpha(u)=u$ imply that
\[
\lim_{k\mapsto0}(\QQ_{2}-\QQ_{1})=0,
\]
and hence
\begin{equation}
\DD_{2}=i(\alpha'_{1}-\alpha'_{2}).\label{eq:D2Linear-1}
\end{equation}
On the other hand, in order to express $\DD_{1}$ and $\DD_{2}$ in
terms of $w'$ and $v_{0}'$ we differentiate formulas (\ref{eq:alpha1})
and (\ref{eq:alpha2}) to obtain 

\begin{eqnarray}
\alpha'_{1} & = & (1-(v_{0}-w/2)k)T+\left(v_{0}'-w'/2\right)N,\label{eq:alpha1deru}\\
\alpha_{2}' & = & (1-(v_{0}+w/2)k)T+\left(v_{0}'+w'/2\right)N,\label{eq:alpha2deru}
\end{eqnarray}
From these equations and formula (\ref{eq:D1Linear}) we obtain that

\[
\Im(\DD_{1})=\arctan\left(\frac{v_{0}'+w'/2}{1-(v_{0}+w/2)k}\right)-\arctan\left(\frac{v_{0}'-w'/2}{1-(v_{0}-w/2)k}\right)
\]
which for $k=0$ becomes
\[
\Im(\DD_{1})|_{k=0}=\arctan\left(v_{0}'+w'/2\right)-\arctan\left(v_{0}'-w'/2\right).
\]
Similarly, from \ref{eq:D2Linear-1} and the above equations for $\alpha'_{1}$
and $\alpha'_{2}$ conclude that
\[
\DD_{2}|_{k=0}=w'.
\]
 Thus 
\[
\DD|_{k=0}=D_{0}\frac{\hbox{Im}(\DD_{1})|_{k=0}}{\Re(\DD_{2})|_{k=0}}=D_{0}\left(\frac{\arctan\left(v_{0}'+w'/2\right)-\arctan\left(v_{0}'-w'/2\right)}{w'}\right),
\]
which is the formula given in \cite{kn:di-projection-diffusion}.

\subsubsection*{An example }

\begin{figure}

\includegraphics[scale=0.12]{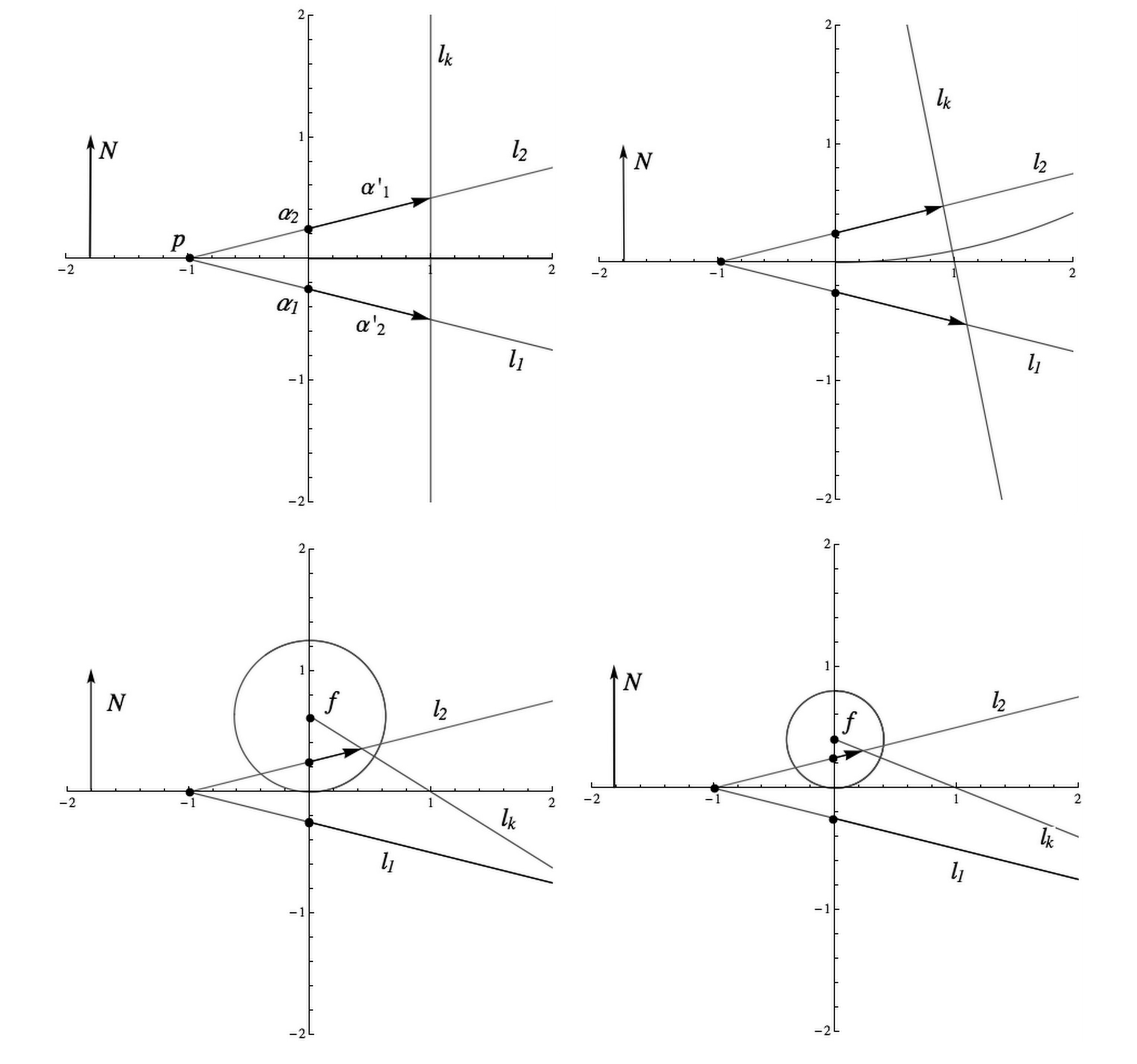}\caption{\label{fig:ExampleProjection}Projection onto a circular curves with
curvatures $k=0,0.2,1.6$ and $2.5$ (left to right and top to bottom)}

\end{figure}

\begin{figure}

\includegraphics[scale=0.12]{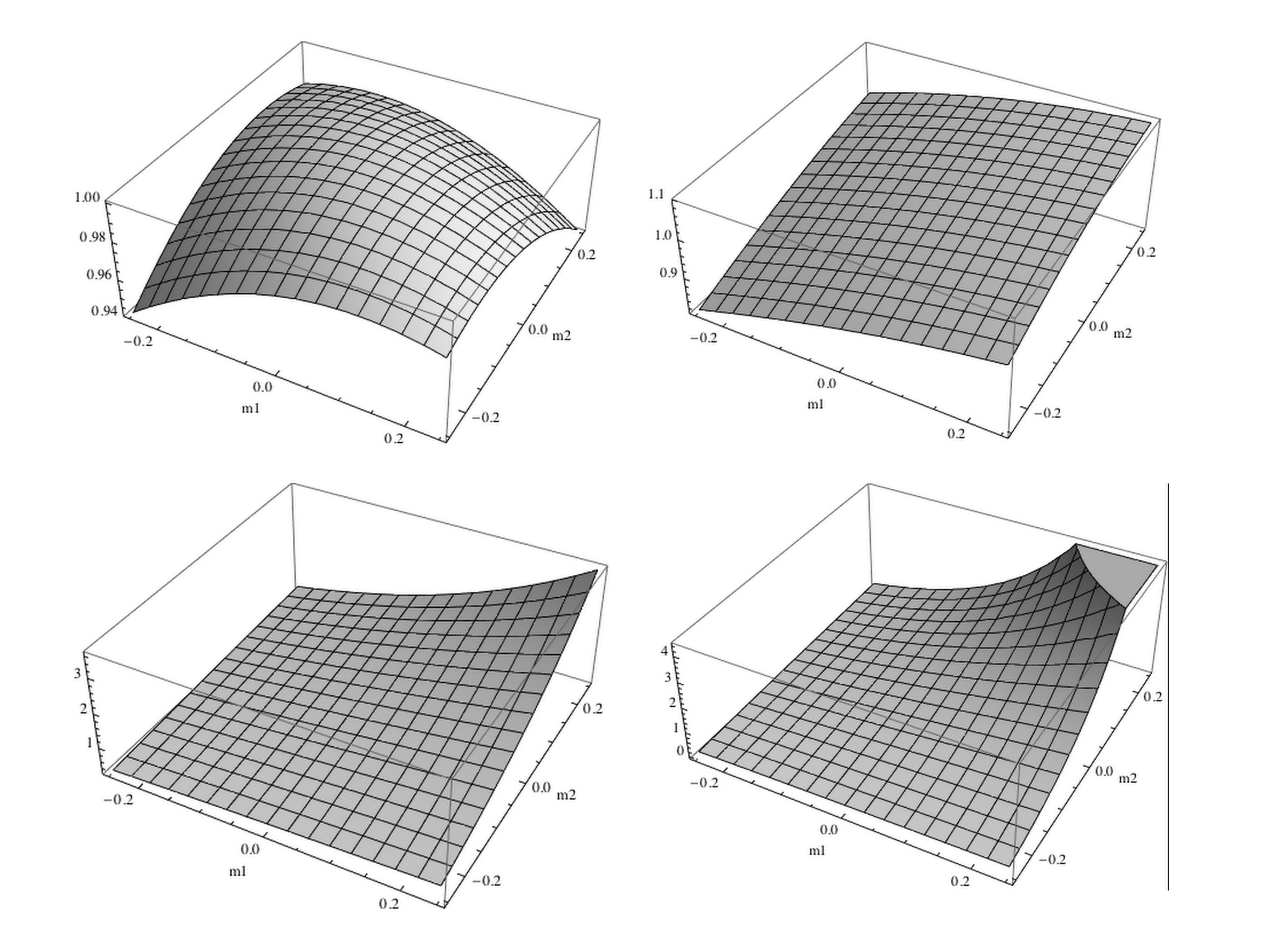}\caption{\label{fig:ExampleEffDiffCoeffLinearCase}Effective diffusion coefficient
as a function of the slopes $m_{1}$ and $m_{2}$ for curvature values
$k=0,0.2,1.6$ and $2.5$ (left to right and top to bottom).}

\end{figure}

To illustrate the behavior of the formula just obtained for $\DD$,
we study the following example. Consider the curve 
\[
\alpha(u)=\frac{i}{k}(1-\exp(iuk))
\]
representing the circle of radius $1/k$ centered at $i/k$. Let $\alpha_{1}$
and $\alpha_{2}$ be the straight lines intersecting at the point
$p=-1$ and with slopes $m_{1}$ and $m_{2}$. We are interested in
visualizing $\DD$ at $u=0$ (see the upper left picture in Figure
\ref{fig:ExampleProjection}) where 
\[
\alpha(0)=0,N(u)=i,\alpha_{1}(0)=m_{1}i\and\alpha_{2}(0)=m_{2}i.
\]
 Let $l_{k}$ be the straight line joining the focal point $f=i/k$
and $1$. If we place the vectors $\alpha'_{1}(0)$ and $\alpha'_{2}(0)$
at the points $\alpha_{1}(0)$ and $\alpha_{2}(0)$ respectively,
they have their end points at the intersections of $l_{k}$ with $l_{1}$
and $l_{2}$. This geometric condition can be written algebraically
as follows
\begin{eqnarray*}
\alpha'_{1}(0) & = & \frac{(1+im_{1})(1-km_{1})}{1+km_{1}},\\
\alpha'_{2}(0) & = & \frac{(1+im_{2})(1-km_{2})}{1+km_{2}}.
\end{eqnarray*}
Substituting the above formulas in our formula for $\DD$, we obtain
$\DD$ as a function of $m_{1},m_{2}$ and $k$. In figure \ref{fig:ExampleProjection}
we show the geometric setting of the above description for the values
$k=0,0.2,1.6$ and $2.5$. Figure \ref{fig:ExampleEffDiffCoeffLinearCase}
we show the resulting effective diffusion coefficient $\DD$ as a
function of $m_{1}$ and $m_{2}$ for the same values of $k$. Observe
that as $k$ gets bigger, the focal point $f$ approaches the line
$l_{2}$ and $\DD$ tends  to become singular.

\subsection{Quadratic case}

\begin{figure}
\includegraphics[scale=0.5]{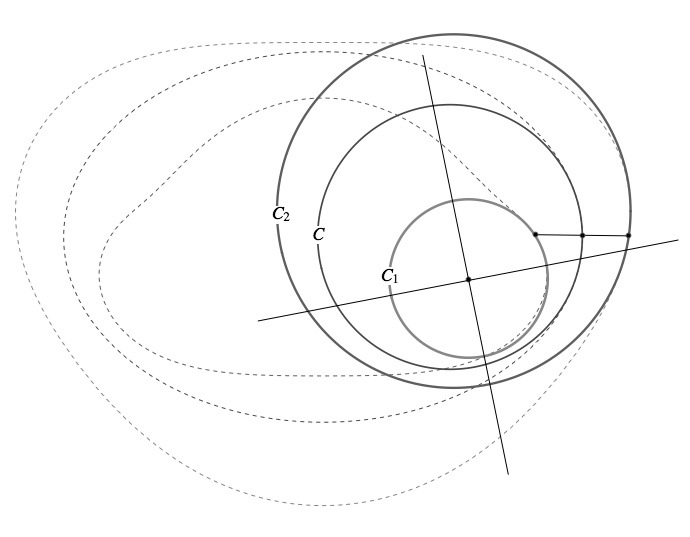}\caption{\label{fig:Aproximation-by-circular}Approximation by circular channel}
\end{figure}

In this case, we approximate the curves $\alpha_{1},\alpha_{2}$ and
$\alpha$ by their corresponding circles of curvature $C_{1},C_{2}$
and $C$ (see Figure \ref{fig:Aproximation-by-circular}). 

We will estimate the effective diffusion coefficient by studying the
effective flux of a stationary solution of the two dimensional diffusion
equation that has no flux across $C_{1}$ and $C_{2}$. As in the
previous sub-section, we will let $\alpha:[-\pi/k,\pi/k]\rightarrow\CC$
be the arc-length parametrization of the circle $C$ of radius $1/k$
and centre $f$ given by
\[
\alpha(u)=f-i\exp(iku)/k,
\]
and the upper and lower walls of our channel $\Omega$ will be circles
$C_{1}$ and $C_{2}$ with radii $1/k_{1},1/k_{2}$ and centers $f_{1},f_{2}$,
respectively. The channel $\Omega$ can be parametrized (at least
locally) by a map $\varphi:[-\pi/k,\pi/k]\times[-1,1]\rightarrow\CC$
of the form

\[
\varphi(u,v)=\alpha(u)+\left(v_{0}(u)+vw(u)/2\right)N(u),
\]
where the functions $v_{0}$ and $w$ are implicitly defined by the
equations
\[
|\alpha_{1}(u)-f_{1}|=1/k_{1}\and|\alpha_{2}-f_{2}|=1/k_{2},
\]
 and $N$, the unit normal field to the circle $C$, is given by
\[
N(u)=\exp(iku).
\]
The middle curve is given by
\[
\alpha_{0}(u)=\varphi(u,0)=\alpha(u)+v_{0}(u)N(u),
\]
and the circles $C_{1}$ and $C_{2}$ have parametrizations
\begin{eqnarray*}
\alpha_{1}(u) & = & \varphi(u,-1)=\alpha(u)+(v_{0}(u)-w(u)/2)N(u),\\
\alpha_{2}(u) & = & \varphi(u,1)=\alpha(u)+(v_{0}(u)+w(u)/2)N(u).
\end{eqnarray*}
In Appendix 2 we show how to find points $q_{1},q_{2}$ in $\CC$
such that the imaginary part of the function (see formulas (\ref{eq:q1Formula})
and (\ref{eq:q2Formula})) 
\[
\PP(z)=\begin{cases}
\II\log\left((z+q_{1})/(z-q_{2}\right) & \hbox{for\,\,}f_{1}\not=f_{2},\\
i\log(z-g) & \hbox{for\,\,}f_{1}=f_{2}=g.
\end{cases}
\]
 is constant on $C_{1}$ and $C_{2}$ (see Figure \ref{fig:Harmonic-Circles}),
and where the coefficient $\II$ is defined by
\[
\II=\begin{cases}
i & \hbox{\,\, if\,\,}C_{1}\cap C_{2}=\emptyset,\\
1 & \hbox{\,\, if\,\,}C_{1}\cap C_{2}\not=\emptyset.
\end{cases}
\]
The real part of $\PP$ is then a stationary solution (i.e. time-independent)
of the two-dimensional diffusion equation that has no flux across
$C_{1}$ or $C_{2}$. We will first consider the case $f_{1}\not=f_{2}$.
Using this $\PP$ in the integral (\ref{eq:rhoIntegral}) we obtain
\[
\rho=Q_{2}\log(\alpha_{2}-q_{1})-Q_{1}\log(\alpha_{1}-q_{1})+Q_{4}\log(\alpha_{2}-q_{2})-Q_{3}\log(\alpha_{1}-q_{2})-R,
\]
where
\begin{eqnarray*}
Q_{1} & = & \frac{k\II(\alpha_{1}-q_{1}-2f)(\alpha_{1}-q_{1})}{2n^{2}},\\
Q_{2} & = & \frac{k\II(\alpha_{2}-q_{1}-2f)(\alpha_{2}-q_{1})}{2n^{2}},\\
Q_{3} & = & \frac{k\II(\alpha_{1}-q_{2}-2f)(\alpha_{1}-q_{2})}{2n^{2}},\\
Q_{4} & = & \frac{k\II(\alpha_{2}-q_{2}-2f)(\alpha_{2}-q_{2})}{2n^{2}},\\
R & = & \frac{k\II(q_{2}-q_{1})(\alpha_{2}-\alpha_{1})}{2n^{2}}.
\end{eqnarray*}
By using the above formula for $\rho$, we obtain 
\[
\DD_{2}=\QQ(q_{2})\log\left(\frac{\alpha_{2}-q_{2}}{\alpha_{1}-q_{2}}\right)-\QQ(q_{1})\log\left(\frac{\alpha_{2}-q_{1}}{\alpha_{1}-q_{1}}\right),
\]
where
\begin{eqnarray*}
\QQ(z) & = & \frac{k\II(\alpha_{1}+z-2f)(\alpha_{1}-z)(\alpha_{2}-f)\alpha'_{2}}{2N^{2}(\alpha_{2}-\alpha_{1})(\alpha_{0}-f)},\\
 &  & -\frac{(\alpha_{1}-f)(\alpha_{2}+z-2f)(\alpha_{2}-z)\alpha'_{1}}{2N^{2}(\alpha_{2}-\alpha_{1})(\alpha_{0}-f)},\\
\RRR(z) & = & \frac{k\II(q_{2}-q_{1})((\alpha_{2}-f)\alpha'_{2}-(\alpha_{1}-f)\alpha'_{1})}{2N^{2}(\alpha_{0}-f)}.
\end{eqnarray*}
For the case $f_{1}=f_{2}=g$ we need to use the function 
\[
\PP(z)=i\log(z-g)
\]
from which we obtain
\[
\rho=R-(Q_{2}\log(\alpha_{2}-g)-Q_{1}\log(\alpha_{1}-g))
\]
where
\begin{eqnarray*}
Q_{1} & = & \frac{ik(\alpha_{1}+g-2f)(\alpha_{1}-g)}{2N^{2}},\\
Q_{2} & = & \frac{ik(\alpha_{2}+g-2f)(\alpha_{2}-g)}{2N^{2}},\\
R & = & \frac{ik(\alpha_{0}+g-2f)(\alpha_{2}-\alpha_{1})}{2N^{2}}.
\end{eqnarray*}
Using this formula for $\rho$ we obtain 
\[
\DD_{2}=\QQ\log\left(\frac{\alpha_{2}-g}{\alpha_{1}-g}\right)-\RRR,
\]
where
\begin{eqnarray}
\QQ & = & \frac{ik((\alpha_{1}+g-2f)(\alpha_{1}-g)(\alpha_{2}-f)\alpha'_{2}}{2N^{2}(\alpha_{2}-\alpha_{1})(\alpha_{0}-f)}\label{eq:D2InfiniteQ}\\
 & - & \frac{(\alpha_{2}+g-2f)(\alpha_{2}-g)(\alpha_{1}-f)\alpha'_{1})}{2N^{2}(\alpha_{2}-\alpha_{1})(\alpha_{0}-f)},\nonumber \\
\RRR & = & \frac{ik(\alpha_{0}+g-2f)((\alpha_{2}-f)\alpha'_{2}-(\alpha_{1}-f)\alpha'_{1})}{2N^{2}(\alpha_{0}-f)}.\label{eq:D2InfiniteR}
\end{eqnarray}

\subsubsection*{Recovering the the case of infinite transversal diffusion rate for
symmetric channels of constant width}

In this case the circles of curvature of $\alpha_{1},\alpha_{2}$
and $\alpha$ have the same center, namely the center of curvature
$f$ of $\alpha$. Hence, we can assume that
\begin{eqnarray*}
\alpha(u) & = & f-i\exp(iku)/k,\\
\alpha_{1}(u) & = & f-i\exp(iku)/k_{1},\\
\alpha_{2}(u) & = & f-i\exp(iku)/k_{2}.
\end{eqnarray*}
If we substitute these formulas in equations \ref{eq:D2InfiniteQ}
and \ref{eq:D2InfiniteR} and let $g=f$ we get
\[
\DD_{1}=i\log(k_{1}/k_{2})\and\DD_{2}=\frac{k^{2}}{2}\left(\frac{1}{k_{1}^{2}}-\frac{1}{k_{2}^{2}}\right).
\]
By using formula (\ref{eq:CurvaturesFront}) we obtain 
\[
k_{1}=\frac{k}{1-kw/2}\and k_{2}=\frac{k}{1+kw/2}
\]
and hence

\[
\DD_{1}=i\log\left(\frac{1+kw/2}{1-kw/2}\right)\and\DD_{2}=kw.
\]
We conclude that
\[
\DD=\frac{D_{0}}{kw}\log\left(\frac{1+kw/2}{1-kw/2}\right),
\]
which is formula (\ref{eq:SymmCstWidth}) for the infinite transversal
diffusion case. In other words, the infinite transversal diffusion
rate formula is actually a second order approximation for symmetric
constant width channels. This results is  reasonable since in this
case the upper and lower curve follows the base curve ``as closely
as possible''.
\[
.
\]

\section{Appendix - A brief review of the differential geometry of plane curves.}

\begin{figure}
\includegraphics[scale=0.5]{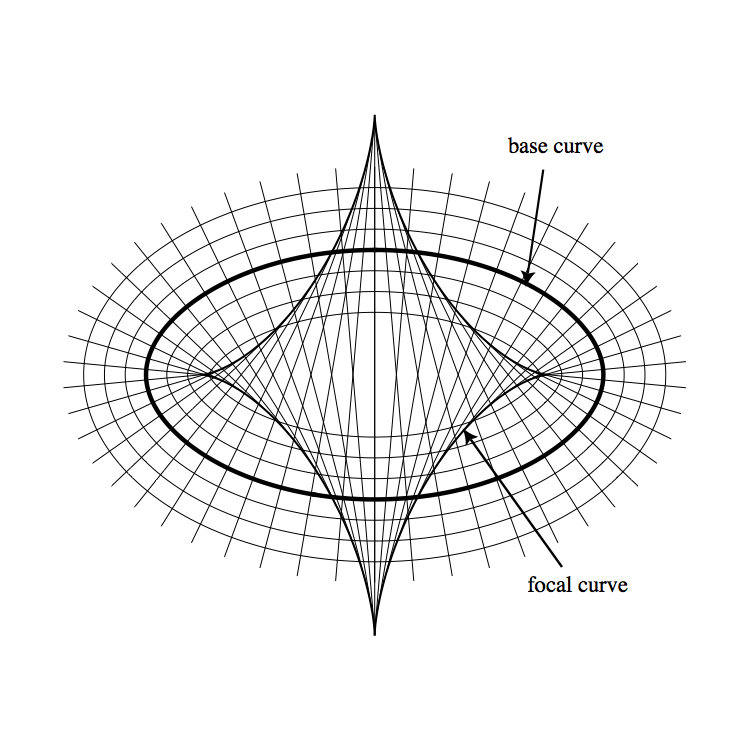}\caption{\label{fig:EllipseFocalCurve}Orthogonal families of curves induced
by an ellipse.}
\end{figure}

In this appendix we review some basic concepts of the differential
geometry of plane curves. The material is standard and can be found
in books such as \cite{kn:spivak2,kn:docarmo}. Consider a smooth
planar curve parametrized by a function $\alpha:[s_{1},s_{2}]\rightarrow\RR^{2}$
where $\alpha(s)=(x(s),y(s))$. The curve is said to have arc-length
parametrization if for all $s$ in the interval $[s_{1},s_{2}]$ we
have that 
\[
\left|\d{\alpha}s\right|=1\where\left|\d{\alpha}s\right|=\sqrt{\left(\d xs\right)^{2}+\left(\d ys\right)^{2}}.
\]
If the above condition holds, then the length of the curve segment
$\alpha([s_{1},s])$ is given by
\[
\hbox{length}(\alpha([s_{1},s])=\int_{s_{1}}^{s}\left|\d{\alpha}s(a)\right|da=s.
\]
In this case we have that 
\[
T=\left(\d xs,\d ys\right)\and N=\left(-\d ys,\d xs\right)
\]
are the unit tangent and unit normal fields to $\alpha$. Since $T$
has unit length its derivative must be a scalar multiple of $N$,
i.e
\[
\d Ts=kN.
\]
The resulting function $k:[s_{1},s_{2}]\rightarrow\RR$ is known as
the \emph{curvature function} of $\alpha$. Similarly
\[
\d Ns=-kT.
\]
For $k(u)\not=0$ the point
\[
f(u)=\alpha(u)+\frac{1}{k(u)}N(u)
\]
is know as the \emph{focal point} of $\alpha$ at $u$, and the scalar
$r(u)=1/k(u)$ is know as the \emph{radius of curvature}. The circle
with centre $f(u)$ and radius $r(u)$ is an approximation of second
order to the curve $\alpha$ at the point $\alpha(u)$. The set of
focal points is known as the \emph{focal curve}. The families of curves
$\alpha_{v}(u)=\alpha(u)+vN(u)$ and $\beta_{u}(v)=\alpha(u)+vN(u)$
form an orthogonal system of curves that becomes degenerate at the
focal curve of $\alpha$ (see Figure \ref{fig:EllipseFocalCurve}).
The curves $\alpha_{v}$ share the same focal curve, and the curvature
function $k_{v}$ of $\alpha_{v}$ can be computed as
\begin{equation}
k_{v}=\frac{k}{1+kv}.\label{eq:CurvaturesFront}
\end{equation}
If the curve $\alpha$ is not parametrized by the arc-length parameter
$s$ but by an arbitrary parameter $t$, we can calculate the curvature
function of $\alpha$ by the following formula
\[
k=\left(\der xt\dder yt-\der yt\dder xt\right)/\left(\left(\der xt\right)^{2}+\left(\der yt\right)^{2}\right)^{3/2}.
\]

\section{Appendix - A complex analytic function whose real part has zero flux
across a pair of arbitrary circles}

The purpose of this appendix is to derive an explicit formula for
a complex analytic function $\PP:\CC\rightarrow\CC$ whose real part
has zero flux across two arbitrary planar circles $C_{1}$ and $C_{2}$.
By the Cauchy-Riemann equations this last condition is equivalent
to the imaginary part of $\PP$ having $C_{1}$ and $C_{2}$ as level
sets. We will denote the centers of $C_{1}$ and $C_{2}$ by $f_{1}$
and $f_{2},$ and their radii by $r_{1}$ and $r_{2}$. The desired
function will have the form

\[
\PP(z)=\II\log\left(\frac{z-q_{2}}{z-q_{1}}\right)=\II\left(\log\left|\frac{z-q_{2}}{z-q_{1}}\right|+i\arg\left(\frac{z-q_{2}}{z-q_{1}}\right)\right),
\]
for adequate values of $q_{1}$ and $q_{2}$ in $\CC$, and where
\[
\II=\begin{cases}
i & \hbox{\,\, if\,\,}C_{1}\cap C_{2}=\emptyset,\\
1 & \hbox{\,\, if\,\,}C_{1}\cap C_{2}\not=\emptyset.
\end{cases}
\]
We will first study the case when both $q_{1}$ and $q_{2}$ are real
numbers with $q_{1}=-q_{2}$ and initially assume that $C_{1}\cap C_{2}=\emptyset$.
We will then explain how to obtain the general case from this one. 

\begin{figure}
\includegraphics[scale=0.5]{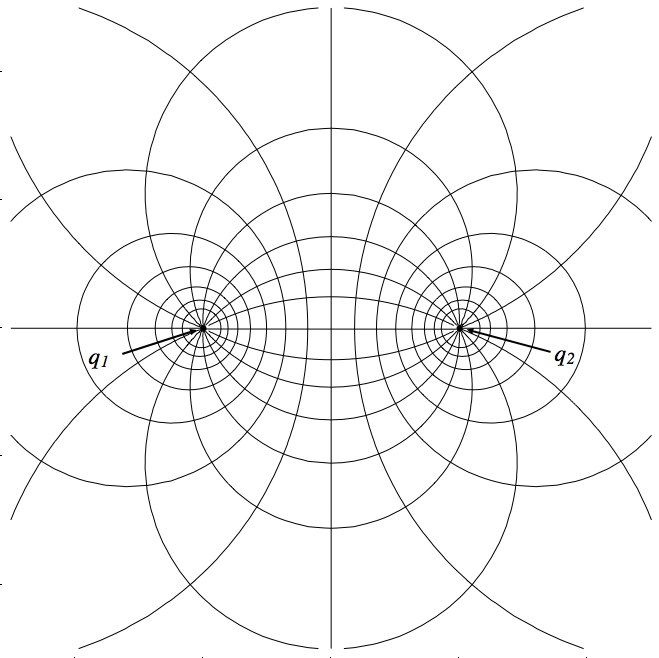}\caption{\label{fig:Steiner-Circles}Steiner Net.}
\end{figure}

\begin{figure}
\includegraphics[scale=0.42]{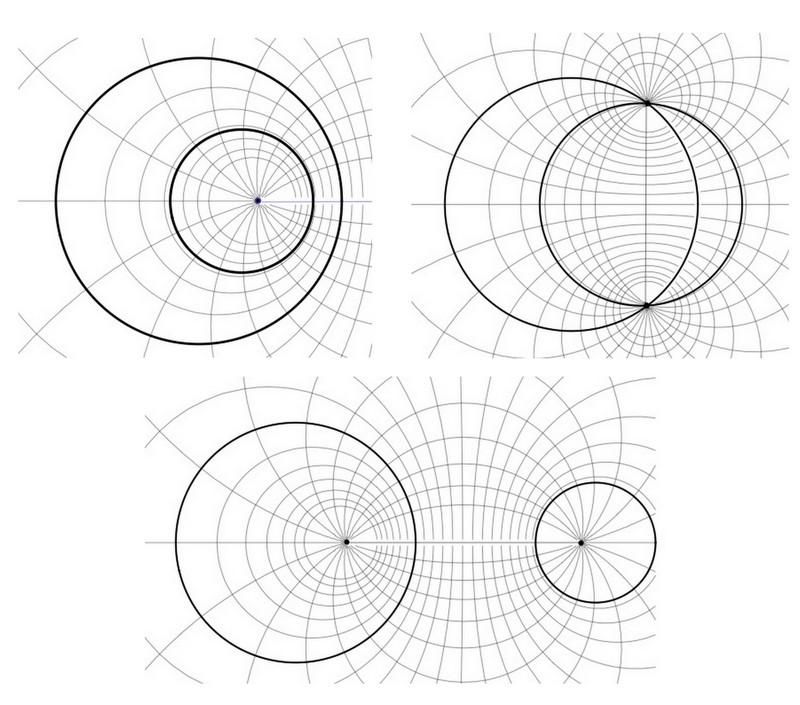}\caption{\label{fig:Harmonic-Circles}Harmonic function having a pair of circles
as level sets}
\end{figure}

For $q\geq0$, we will write $q_{1}=q$, $q_{2}=-q$ so that 
\[
\PP(z)=i\log\left(\frac{z+q}{z-q}\right).
\]
The solutions of the equations 
\begin{eqnarray}
\Im(\PP(z)) & = & \log\left(\alpha\right),\label{eq:AppCircles}\\
\Re(\PP(z)) & = & \theta,\label{eq:AppCirclesOrth}
\end{eqnarray}
define a net of circles known as the Steiner net (see Figure \ref{fig:Steiner-Circles}).
The radius $r$ and center $c$ of a circle defined by equation \ref{eq:AppCircles}
are given by
\[
r=\frac{2q\alpha}{|\alpha^{2}-1|}\and c=q\left(\frac{\alpha^{2}+1}{\alpha^{2}-1}\right).
\]
We can solve these equations for $\alpha$ and $f$ to obtain 
\[
\alpha=\frac{\sqrt{q^{2}+r^{2}}\pm q}{r}\and c=\pm\sqrt{q^{2}+r^{2}},
\]
where the choice of signs must be the same in both formulas, and the
sign determines if the corresponding circle of radius $r$ lies on
the positive or  negative side of the real axis. Hence, for any $q$
the circles radii $r_{1}$ and $r_{2}$ correspond to the values 
\[
\alpha_{1}=\frac{\sqrt{q^{2}+r_{1}^{2}}\pm q}{r_{1}}\and\alpha_{2}=\frac{\sqrt{q^{2}+r_{2}^{2}}\pm q}{r_{2}}.
\]
We can fix the distance $d=|c_{2}-c_{1}|$ between the centers of
the circles by letting 
\begin{equation}
q=\frac{1}{2d}\sqrt{\left(d^{2}-(r_{1}+r_{2})^{2}\right)\left(d^{2}-(r_{2}-r_{1})^{2}\right)},\label{eq:qValue}
\end{equation}
Having chosen this $q$ we can recover the centers of the circles
by the formulas
\[
c_{1}=\pm\sqrt{q^{2}+r_{1}^{2}}\and c_{2}=\pm\sqrt{q^{2}+r_{2}^{2}}.
\]
We will choose $c_{1}$ to be positive. If $c_{2}$ is negative the
circles are outside each other, and if $c_{2}$ is positive then one
circle is inside of the other. It turns out that for $q$ given by
(\ref{eq:qValue}) a small correction to our formula for $\PP$ will
even work for the case when $q$ is complex. This last condition corresponds
to the case when $C_{1}$ and $C_{2}$ intersect (see Figure \ref{fig:Harmonic-Circles}).
In order to introduce the needed correction, we will need the functions
\[
\II=\begin{cases}
i & \hbox{if\,\,\,}q\hbox{\, is\, real},\\
1 & \hbox{if\,\,\,}q\hbox{\, is\, complex}.
\end{cases}
\]
and
\[
\JJ=\begin{cases}
1 & \hbox{if\,\,\,}(\II=1\and d<c_{1})\hbox{\,\, or\,\,\,}(\II=i\and d<r_{1}+r_{2})\\
-1 & \hbox{otherwise}
\end{cases}
\]
Our final formula, which deals with the case of general $q_{1}$ and
$q_{2}$ can be obtained by using the transformation

\[
T(z)=\frac{(c_{2}-c_{1})z+c_{1}f_{2}-c_{2}f_{1}}{f_{2}-f_{1}}
\]
that satisfies $T(f_{1})=c_{1},T(f_{2})=c_{2}$. For $d=|f_{2}-f_{1}|$
and $q$ given by equation (\ref{eq:qValue}) the function with the
desired properties is
\[
\PP(z)=\II\log\left(\frac{z-q_{1}}{z-q_{2}}\right),
\]
where

\[
c_{1}=\sqrt{q^{2}+r_{1}^{2}},c_{2}=\JJ\sqrt{q^{2}+r_{2}^{2}}.
\]
and 
\begin{eqnarray}
q_{1} & = & \frac{f_{1}c_{2}-f_{2}c_{1}+q(f_{2}-f_{1})}{c_{2}-c_{1}},\label{eq:q1Formula}\\
q_{2} & = & \frac{f_{1}c_{2}-f_{2}c_{1}-q(f_{2}-f_{1})}{c_{2}-c_{1}}.\label{eq:q2Formula}
\end{eqnarray}
For $f_{1}=f_{2}$ (i.e $d=0$) the function $\PP$ is not well defined.
We correct this by letting 
\[
\PP(z)=\begin{cases}
\II\log\left((z-q_{2})/(z-q_{1})\right) & \hbox{for\,\,}f_{1}\not=f_{2},\\
i\log(z-g) & \hbox{for\,\,}f_{1}=f_{2}=g.
\end{cases}
\]

\section{Acknowledgments}

The first author would like to thank Marco-Vinicio Vazquez Gonzalez
for many useful conversations. The second author would like to thank
the International Centre for Theoretical Physics for its support and
hospitality.\bibliographystyle{plain}
\bibliography{myBib}

\end{document}